# A Macroscopic Approach to Creating Exotic Matter

Charles T. Ridgely
charles@ridgely.ws

Herein the Casimir effect is used to present a simple macroscopic view on creating exotic matter. The energy arising between two nearly perfectly conducting parallel plates is shown to become increasingly negative as the plate separation is reduced. It is proposed that the Casimir energy appears increasingly negative simply because the vacuum electromagnetic zero-point field performs positive work in pushing the plates together, transforming field energy into kinetic energy of the plates. Next, the inertial properties of exotic matter are considered. The parallel plates of the Casimir system are replaced with an enclosed cavity of identical dimensions that is subjected to an external force. It is found that zero-point radiation exerts an inertial force on the cavity in opposition to the external force. This ultimately leads to the conclusion that the inertial properties of exotic matter are identical to the inertial properties of ordinary matter.

## 1. Introduction

According to a recent proposal by Miguel Alcubierre [1], general relativity admits metric solutions that predict the possibility of superluminal space travel. In essence, these metrics imply that a spaceship can be propelled through space by expanding space-time behind the ship while, at the same time, contracting space-time in front of the ship. The ship then surfs on this space-time deformation, acquiring very high velocity relative to the rest of the universe. Even more surprising is that these 'warp drive' metrics predict that such travel can be carried out without the usual time dilation effects predicted by special relativity. That is, proper time experienced by a space traveler, using an Alcubierre warp drive, is identical to the time experienced by observers remaining on the Earth.

As with worm holes, warp drives require tremendous quantities of negative energy, or exotic matter [1]-[5]. Although there are several conceptual difficulties associated with warp drive, the energy requirement alone can easily be taken as sufficient grounds for us to dismiss the feasibility of warp drive. More encouraging, however, is that modified forms of the Alcubierre metric have been proposed, requiring more reasonable quantities of exotic matter [4]-[5]. We can only hope that as more research is conducted, warp drive theory will become increasingly tenable.

Of course, there is one serious obstacle to our actually engineering a warp drive system of *any* size: the need for exotic matter. Before we can set up a physically realizable warp drive, we must have a working knowledge of how to generate exotic matter. As is well known, exotic matter is forbidden classically, but is permitted on the quantum level under special circumstances. When these special circumstances are applied to warp drive theory, we are left with a warp bubble requiring massive amounts of exotic matter and having walls whose thickness is on the order of the Planck length [2]-[3]. But even with such constraints, we are still left with the question of how to obtain exotic matter, regardless of quantity. It is an objective of the present analysis to provide a simple macroscopic concept for creating exotic matter, while leaving the question of quantity for future development.

One well-known example in which negative energy produces observable forces is the Casimir effect [1], [4]. The Casimir effect is an attractive force that arises between two nearly perfectly conducting parallel plates when they are placed in close proximity, as shown in Fig. 1. In essence, the plates block out some of the vacuum zero-point electromagnetic radiation that would otherwise reside in the volume between the plates, were the plates removed [6]-[7]. This gives rise to a small energy difference between the plates relative to the region outside the plates:

$$U(r) = -\frac{\pi^2}{720}\frac{\hbar c L^2}{r^3} \quad (1)$$

where $L^2$ is the surface area of each plate, and $r$ is the plate separation. Differentiating Eq. (1) with respect to the plate separation then gives the force on the plates:

$$F(r) = -\frac{\partial U(r)}{\partial r} = -\frac{\pi^2}{240}\frac{\hbar c L^2}{r^4} \quad (2)$$

This attractive force, tending to push the plates together, has been experimentally verified to within 5% of its theoretical value [7].

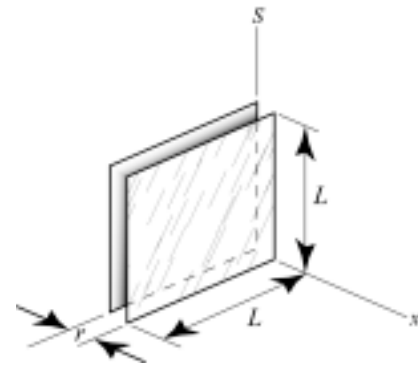

**Figure 1.** Two nearly perfectly conducting plates of area $L^2$ are placed in parallel, separated by distance $r << L$.

While the Casimir effect is typically reserved for discussions involving the quantum world, there may in fact be a way to picture the negative energy-content of the Casimir system from a macroscopic point of view. This can be done by taking a macroscopic view of the vacuum electromagnetic zero-point field (ZPF). With this approach, the ZPF is viewed as a real Lorentz-invariant radiation field that is essentially uniform throughout all of space. So long as the ZPF is isotropic, no direct observation of the ZPF can be made; but if the ZPF becomes locally anisotropic, such as in the Casimir system, then direct observations can then indeed be carried out [7]-[8].

The next Section deals with the change in Casimir energy that arises when the plate separation decreases. It is shown that the ZPF performs positive work in pushing the plates together. The resulting drop in field energy appears increasingly negative with respect to the uniform energy content outside the plates [8]-[9]. Based on this, it is proposed that whenever the ZPF performs positive work in a localized region of space-time, a drop in zero-point field energy occurs, resulting in a negative energy density relative to the rest of space-time.

Section 3 is devoted to acquiring greater insight into the inertial properties of exotic matter. Applying the law of inertia of energy [10]-[11] to the Casimir energy leads to a negative inertial mass. Using this, it is shown that exotic matter possesses inertial properties identical to those of ordinary matter.

## 2. Negative Energy due to Work Performed by Zero-Point Field

Consider the energy between two parallel plates when their separation is first $r_1$ and then later $r_2$, where $r_1 > r_2$. According to Eq. (1), when the plate separation is $r_1$ and $r_2$ the energy between the plates is, respectively

$$U_1 = -\frac{\pi^2}{720}\frac{\hbar c L^2}{r_1^3}, \quad U_2 = -\frac{\pi^2}{720}\frac{\hbar c L^2}{r_2^3} \quad (3a,b)$$

As pointed out in the Introduction, the force between the plates is attractive, tending to push the plates together. Suppose the force pushes the plates from separation $r_1$ to separation $r_2$. The change in energy between the plates is then

$$\Delta U = U_2 - U_1 = -\frac{\pi^2}{720}\hbar c L^2 \left(\frac{1}{r_2^3} - \frac{1}{r_1^3}\right) \quad (4)$$

This implies that the Casimir energy becomes increasingly negative as the plate separation is reduced. The work performed in order to bring about this change in energy is

$$W_{12} = \frac{\pi^2}{720}\hbar c L^2 \left(\frac{1}{r_2^3} - \frac{1}{r_1^3}\right) \quad (5)$$

This is positive work performed by the ZPF in pushing the plates from $r_1$ to $r_2$, transforming electromagnetic field energy into kinetic energy of the plates. Since the energy content of the ZPF is initially uniform everywhere [8]-[9], this drop in field energy manifests as a negative energy density between the plates, relative to the energy density of the region outside the plates.

Of particular interest is the case in which the initial separation of the parallel plates is very large. Assuming that no other forces act on the plates, the ZPF should still exert a force on the plates, no matter how minute that force may be. The work performed by that force can be expressed simply by taking the limit of Eq. (5) as $r_1$ tends to infinity:

$$W_\infty = \lim_{r_1 \to \infty}(W_{12}) \quad (6)$$

Carrying this out leads directly to

$$W_\infty = \frac{\pi^2}{720}\frac{\hbar c L^2}{r^3} \quad (7)$$

in which the subscript on $r_2$ has been dropped for simplicity. This is the work performed by the ZPF in pushing two parallel plates to a separation $r$ when the plates are initially very far apart. The potential energy associated with Eq. (7) is then

$$U_\infty = -\frac{\pi^2}{720}\frac{\hbar c L^2}{r^3} \quad (8)$$

which is identical to the original expression for the Casimir energy, given by Eq. (1). So then, we have come full circle, but with the clear assertion that the negative energy density of the Casimir system arises because the ZPF performs positive work in a localized region of space-time. More specifically, we propose that in any instance in which the ZPF performs positive work in a localized region of space-time, a drop in vacuum electromagnetic zero-point field energy occurs, resulting in a negative energy density in that region relative to the rest of space-time [9].

## 3. Inertial Properties of Exotic Matter

Taking a strict view of the relationship between energy and mass [10], we might suppose that such a region exhibits an inertial mass in accordance with the law of inertia of energy, $m = U_\infty/c^2$. Applying this expression to Eq. (8) leads directly to

$$m = -\frac{\pi^2}{720}\frac{\hbar L^2}{c r^3} \quad (9)$$

This implies that the mass-energy between the plates of the Casimir system is negative. At first sight, we might expect this small quantity of exotic matter should reduce the overall inertial mass of the Casimir system. As we shall see, however, we have no good reason to presuppose that exotic matter behaves in any way differently than does ordinary matter.

According to previous analyses [10]-[12], the inertial properties of energy cannot be adequately explained on the basis of inertial mass alone; the relativistic nature of space-time must also be taken into account. This was demonstrated by considering an ideal cavity of length $r$, containing monochromatic radiation of frequency $v_0$ that is subjected to an external force [10]. It was found that when an external force is applied to such a cavity, radiation pressure exerts an inertial resistance force on the cavity, given by

$$\bar{F} = \frac{\bar{N}h}{r}(v_A - v_B)\hat{n} \quad (10)$$

where $\bar{N}$ is the average number of photons in the cavity, $v_A$ and $v_B$ are the frequencies of radiation measured by co-moving observers at walls *A* and *B* of the cavity, respectively, and $\hat{n}$ is a unit vector in the direction of the cavity's acceleration. This expression for the force was used to show that the inertial resistance of the cavity arises as a direct manifestation of space-time anisotropy within the cavity [10].

The plan here is to use Eq. (10) to gain greater insight into the inertial properties of exotic matter. To do this, the parallel plates of the Casimir system are replaced with a cavity of identical dimensions, namely one with $r << L$, as shown in Fig. 2. Also, the zero-point radiation under consideration here is not monochromatic, and thus the average indicated by $\bar{N}$ must be eliminated in favor of a sum over all frequency modes within the cavity. Carrying this out, Eq. (10) can be recast in the form

$$\bar{F} = \frac{\hbar}{r}\left[\left(\sum_i^N \omega_i\right)_A - \left(\sum_i^N \omega_i\right)_B\right]\hat{n} \quad (11)$$

where *N* is the number of modes within the cavity [13]. For the case of a cavity in thermal equilibrium with its surroundings, the number of photons observed at each wall within the cavity is roughly the same. Using this to our advantage, Eq. (11) can then be written as

$$\bar{F} = \frac{\hbar}{r}\sum_i^N (\omega_A - \omega_B)_i \hat{n} \quad (12)$$

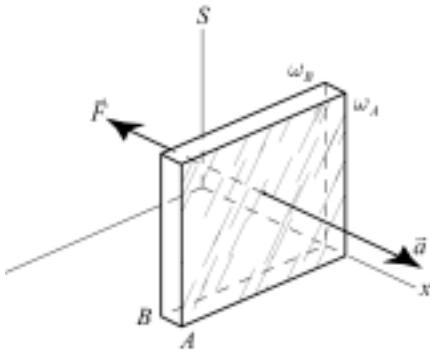

**Figure 2.** An enclosed cavity of identical dimensions as the Casimir system accelerates uniformly under the action of an external force. Space-time anisotropy, arising due to acceleration, gives rise to a frequency-shift in the zero-point radiation observed at walls *A* and *B* within the cavity: $\omega_A - \omega_B \neq 0$. Zero-point radiation pressure exerts a net force $\bar{F}$ on the cavity in opposition to the externally applied force.

This expression implies that when an external force is applied to the cavity, each frequency mode within the cavity exerts a net force on the cavity, given by

$$\bar{F}_i = \frac{\hbar}{r}(\omega_A - \omega_B)_i \hat{n} \quad (13)$$

For the case of a uniformly accelerating cavity, as shown in Fig. 2, anisotropy of space-time gives rise to an observable Doppler-shift within the cavity [10]: $\omega_A - \omega_B \neq 0$. Observers residing in the co-moving reference frame (CMRF) of the cavity find that zero-point radiation arriving at wall *B* is blue-shifted relative to radiation observed at wall *A*. If the frequencies observed at wall *A* are $\omega_{Ai} = \omega_{0i}$, then the frequencies at wall *B* are

$$\omega_{Bi} = \omega_{0i}\sqrt{\frac{1 + V_R/c}{1 - V_R/c}} \quad (14)$$

where $V_R$ is the relative velocity of wall *B* gained during the time in which photons travel from wall *A* to wall *B*. Using these frequencies, observers moving with the cavity recast Eq. (12) in the form

$$\bar{F} = \frac{\hbar}{r}\sum_i^N \left(\omega_0 - \omega_0\sqrt{\frac{1 + V_R/c}{1 - V_R/c}}\right)_i \hat{n} \quad (15)$$

Rearranging a bit, this expression becomes

$$\bar{F} = \sum_i^N (\hbar\omega_0)_i \left[\frac{1}{r}\left(1 - \sqrt{\frac{1 + V_R/c}{1 - V_R/c}}\right)\right]\hat{n} \quad (16)$$

To further simplify this expression, we notice that for the case of a very small cavity undergoing Newtonian acceleration, we can use the approximation [10]

$$\frac{1}{r}\left(1 - \sqrt{\frac{1 + V_R/c}{1 - V_R/c}}\right) \approx -\bar{\nabla}\left(\frac{dt}{d\tau}\right) \quad (17)$$

where $d\tau$ is an interval of proper time experienced by observers moving with the cavity, $dt$ is an interval of coordinate time experienced by a co-moving observer whose coordinate origin is momentarily coincident with that of the accelerating system at a time $t = 0$, and the minus sign arises because the gradient is in the direction of the cavity's acceleration [10]. Substituting Eq. (17) into Eq. (16) then gives the force as

$$\bar{F} = -\sum_i^N (\hbar\omega_0)_i \bar{\nabla}\left(\frac{dt}{d\tau}\right) \quad (18)$$

As a final consideration, we notice that the summation simply expresses the total zero-point energy within the cavity:

$$E = \sum_{i}^{N} (\hbar \omega_0)_i \quad (19)$$

At this point, it must be stressed that although the energy density within the cavity is negative with respect to exterior regions, individual photons inside the cavity have their usual positive energies, $E_i = \hbar \omega_i$. The cavity simply blocks out those photons having wavelengths longer than the distance between the walls of the cavity [7],[13].

Using Eq. (19) in Eq. (18), the force on the cavity then becomes

$$\vec{F} = -E \vec{\nabla} \left( \frac{dt}{d\tau} \right) \quad (20)$$

According to this expression, zero-point radiation within the cavity exerts an inertial force on the cavity that acts in opposition to the externally applied force, just as does ordinary, positive radiation [10], [14]. Thus, the inertial properties of a region having a negative energy density are identical to those of a region having an equivalent positive energy density. Previous analyses have shown that all forms of positive energy resist changes to their states of motion, and that the source of such resistance is entirely due to space-time anisotropy arising within accelerating systems of reference [10]-[12]. Equation (20) implies that the same holds true for negative forms of energy as well. We can conclude, therefore, that the inertial properties of exotic matter are the same as the inertial properties of ordinary matter. Both forms of matter resist changes to their states of motion when acted upon by an external force.

Returning to the Casimir system, it is now easy to see that the inertial mass predicted by Eq. (9) will contribute positively to the overall inertia of the Casimir system rather than reducing it, as the minus sign initially suggested. While the minus sign in Eq. (9) certainly implies that the mass is exotic, having its origin in negative energy, the minus sign tells us nothing about the inertial properties of exotic matter.

## 4. Discussion

Current theory places serious restrictions on the existence of exotic matter. According to recent analyses, such restrictions set the thickness of the walls of an Alcubierre warp bubble on the order of the Planck length, while requiring an enormous quantity of exotic matter [2]-[3]. In the present analysis, a simple macroscopic concept for creating exotic matter and predicting its inertial properties has been proposed.

Of course, the validity of the interpretation set forth herein ultimately hinges on whether or not the ZPF is comprised of real radiation, and on the amount of work that can be performed thereby. Judging from the literature on the subject, whether or not the ZPF is real seems to be an ongoing debate [7]-[9]. Assuming that the ZPF is indeed real and capable of performing electromagnetic work [9], we propose that exotic matter is created whenever the ZPF performs positive work in a localized region of space.

As a final thought, it is interesting to speculate on how the notions put forth herein might be applied to warp drive theory, mentioned in the Introduction. From a hypothetical standpoint, we can imagine a spaceship, residing in flat space-time, which carries out some sort of internal process that alters the ZPF outside the ship [15]. If, through this process, the ship can coax the ZPF into performing substantial electromagnetic work, then the ship can use the ZPF to generate the exotic matter required for warp drive. In essence, we imagine a spaceship that manipulates the ZPF in order to generate stress-energy outside the ship. With the right configuration, this stress-energy can be used to alter the geometry of space-time outside the ship, forming an Alcubierre-type geometry, or some more efficient variation thereof [1], [4]-[5]. Assuming that all goes as planned, the spaceship should then be propelled by the resulting space-time geometry, acquiring very high velocity relative to the rest of the Universe. One noteworthy feature of this model is that there is no need for the spaceship to carry a quantity of exotic matter on board for use as a fuel source. Rather, such a spaceship would generate exotic matter outside the ship as it moves through space.